\documentclass[11pt,a4paper]{article}
\pdfoutput=1 

\usepackage{jcappub}
\usepackage{mathtools}

\newcommand{\eulere}{\mathrm{e}}

 \usepackage{graphicx}
\usepackage{subcaption}

\title{Consistent early and late time cosmology from the RG flow of gravity}

\newcommand{\addressRoma}{Dipartimento di Fisica, Universit\`a di Roma ``La Sapienza'', P.le A. Moro 2, 00185 Roma, Italy}
\newcommand{\addressRadboud}{Radboud University, Institute for Mathematics, Astrophysics and Particle Physics, Heyendaalseweg 135, NL-6525 AJ Nijmegen, The Netherlands}
 
\author[a,b]{Giulia Gubitosi}
\author[a]{Robin Ooijer}
\author[a]{Chris Ripken}
\author[a]{Frank Saueressig}

\affiliation[a]{\addressRadboud}
\affiliation[b]{\addressRoma}

\emailAdd{g.gubitosi@science.ru.nl}
\emailAdd{robin.ooijer@student.ru.nl}
\emailAdd{aripken@science.ru.nl}
\emailAdd{f.saueressig@science.ru.nl}

\abstract{
We investigate the compatibility of cosmological constraints on inflation and the cosmological constant with the asymptotic safety scenario of quantum gravity. The effective action is taken to be of $f(R)$ form, truncated to second order. The flow generated by the Functional Renormalisation Group Equation is analysed and it is found to allow for trajectories that are compatible with the observational constraints on the parameters of the action, both at early and late cosmological times. 
In particular, the gravitational effective dynamics generated in the trans-Planckian regime flows to Starobinsky inflation at early times and to standard Einstein Gravity with a cosmological constant at late times. Moreover, the cosmological constant acquires an energy dependence at $10^{-2}$ eV, increasing from its current value of $10^{-66} \,\text{eV}^{2}$ on Hubble scale to a value of $10^{30}\, \text{eV}^{2}$ at inflation scale.
}
\begin{document}

\maketitle

\section{Introduction}
\label{sec:introduction}
Observations of supernova explosions \cite{Riess:1998cb,Perlmutter:1998np} suggest that the universe is currently undergoing a phase of accelerated expansion. Moreover, there are strong indications that the universe also went through an early period of accelerated expansion \cite{Ade:2015lrj} called inflation. This dynamics may be recovered by introducing an inflaton field driving the dynamics of the early universe and a positive cosmological constant triggering the accelerated expansion at late times.
In this paper, we explore the possibility that physics at trans-Planckian energies may set the seed for the observed effective dynamics of gravity below the Planck scale.

Our starting point is power-law $f(R)$-gravity, truncated at the second order in the Ricci scalar $R$,
\begin{equation}
			S[g]
	=
		\frac{1}{16\pi G}\int	\operatorname{\mathrm{d}}^4 x	\sqrt{-g}	\left(
			2\Lambda
		-	R
		+	B	R^2
		\right)
	\text{.}\label{eq:action}
\end{equation}
At early cosmological times, the higher-order curvature term gives rise to inflation through the classical equations of motion \cite{Starobinsky:1980te, Mukhanov:1981xt, Starobinsky:1983zz}. Therefore, inflation can be seen as a purely gravitational effect, where the inflaton field is simply the additional gravitational scalar degree of freedom described by the $f(R)$ action. The (constant) lowest-order term plays the role of the cosmological constant, and drives the late-time accelerated expansion. The action is
parameterized by three couplings, 
the cosmological constant $\Lambda$, Newton coupling $G$, and the $R^2$-coupling $B$. 
Cosmological observations impose restrictions on these parameters. Each of them is  measured over a different distance scale, as explained in detail in section~\ref{sec:observational_constraints} and summarized in table~\ref{tab:constraints1}. In this paper we investigate whether the measured values of the parameters are consistent with the asymptotic safety scenario of quantum gravity \cite{Niedermaier:2006wt,Codello:2008vh,Litim:2011cp,Reuter:2012id,Nagy:2012ef,Eichhorn:2017egq,Percacci:2017fkn}.\footnote{Cosmological implications of Asymptotic Safety have received considerable attention \cite{Bonanno:2001hi,Bonanno:2001xi,Guberina:2002wt,Babic:2004ev,Reuter:2005kb,Bonanno:2005mt,Bonanno:2007wg,Weinberg:2009wa,Bonanno:2010bt,Contillo:2011ag,Cai:2011kd,Reuter:2003ca,Tronconi:2017wps,Cai:2013caa,Copeland:2013vva,Saltas:2015vsc,Bonanno:2012jy,Bonanno:2015fga,Bonanno:2018gck,DOdorico:2015jtl}, also see \cite{Bonanno:2017pkg} for an up-to-date review. These works incorporate the leading quantum gravity effects by a so-called renormalization group improvement procedure which identifies the energy scale $k$ with a physical quantity. In this work, we pioneer a different path: instead of generating an effective dynamics via renormalization group improvement, we construct the effective action of the theory valid at the corresponding energy scale by solving the underlying renormalization group equations.}
\begin{table}[t]
	\renewcommand{\arraystretch}{1.5}
	\centering
	\begin{tabular}{lp{.02\textwidth}l}
		Energy scale (eV)	&&	Constraint
	\\	\hline
		$ k_{\text{infl}} = 10^{22}$	&& $B =  -1.7 \times 10^{-46}$ eV$^{-2}$
	\\
		$ k_{\text{lab}} =  10^{-5}$	&& $G = 6.7 \times 10^{-57}$ eV$^{-2}$
	\\
		$ k_{\text{Hub}} = 10^{-33}$	&& $\Lambda= 4 \times 10^{-66}$ eV$^{2}$
	\\	\hline
	\end{tabular}
	\caption{Observational constraints on the parameters of the action \eqref{eq:action}. For each parameter we indicate the energy scale corresponding to the distance over which the measurement is performed. These are the scale of inflation $k_{\text{infl}}$, the laboratory scale $k_{\text{lab}}$, and  the Hubble scale $k_{\text{Hub}}$.}
	\label{tab:constraints1}
\end{table}

The key idea of our analysis is that couplings like the ones appearing in the action~\eqref{eq:action} acquire an energy dependence if gravity is promoted to a quantum field theory. This energy dependence is encoded in the renormalization group (RG) flow of the theory and captured by its $\beta$-functions. A consistent description of gravity valid on all scales may then be obtained along Weinberg's asymptotic safety conjecture \cite{Weinberg:1980gg}. In this scenario the gravitational interactions at trans-Planckian energy are controlled by a non-Gaussian fixed point (NGFP) of the gravitational RG flow. Solutions of the RG equations which are dragged into this fixed point for increasing energy are termed asymptotically safe. The asymptotic safety condition then places restrictions on the admissible values of the couplings and equips the construction with predictive power.
Starting from the seminal work by Reuter \cite{Reuter:1996cp}, there is substantial evidence that such a UV fixed point for gravity exists \cite{Niedermaier:2006wt,Codello:2008vh,Litim:2011cp,Reuter:2012id,Nagy:2012ef,Eichhorn:2017egq,Percacci:2017fkn}. In particular, this fixed point persists in the presence of the  Goroff-Sagnotti two-loop counterterm \cite{Gies:2016con}. 

Contact to the constraints listed in table~\ref{tab:constraints1} is then made by following the RG flow emanating from the NGFP towards low energies. In this work this flow is constructed based on the (Euclidean) action\footnote{The relation between gravitational RG flows obtained from an Euclidean and Lorentzian setting has been studied in \cite{Manrique:2011jc} and it was shown that the two settings lead to qualitatively identical phase diagrams. In the sequel we assume that this result also holds at the level of the $R^2$-type actions \eqref{eq:action} and \eqref{eq:RGaction}.}
\begin{equation}
			S_{k}[g]
	=
			\frac{1}{16\pi G_{k}}\int	\operatorname{\mathrm{d}}^4 x	\sqrt{g}	\left(
				2\Lambda_{k}
			-	R
			+	B_{k}	R^2
			\right)
	\text{,}\label{eq:RGaction}
\end{equation}
where the subscript $k$ implies that the couplings depend on the energy scale $k$. Within this setting, an asymptotically safe RG trajectory describing Nature (including a quantum gravity induced inflationary phase) should then meet the requirements summarized in table~\ref{tab:constraints}. The existence of a RG trajectory that meets all these constraints is highly nontrivial and constitutes the main result of this work.
\begin{table}[t]
	\renewcommand{\arraystretch}{1.5}
	\centering
	\begin{tabular}{lp{.02\textwidth}l}
		Energy scale (eV)	&&	RG constraint
	\\	\hline
		$k \gg M_P = 2.4 \times 10^{27}$ &&	NGFP
	\\
		$k	\simeq k_{\text{infl}} = 10^{22}$	&& $B_k \simeq B_{\text{infl}} = -1.7 \times 10^{-46}$ eV$^{-2}$
	\\
		$k \simeq k_{\text{lab}} =  10^{-5}$	&& $G_k \simeq G = 6.7 \times 10^{-57}$ eV$^{-2}$
	\\
		$k \simeq k_{\text{Hub}} = 10^{-33}$	&& $\Lambda_k \simeq \Lambda = 4 \times 10^{-66}$ eV$^{2}$
	\\	\hline
	\end{tabular}
	\caption{Constraints on the $f(R)$ action parameters in the RG framework at various scales. The observational constraints are the same as in table~\ref{tab:constraints1}. Moreover, we require that beyond the (reduced) Planck scale $M_P \equiv (8\pi G)^{-1/2}$, the RG flows towards a non-Gaussian fixed point (NGFP).}
	\label{tab:constraints}
\end{table}

In parallel to the present work, ref.\ \cite{Liu:2018aa} preformed a similar cosmological study based on the refined Starobinsky model
\begin{equation}\label{refinedStarobinsky}
S = \int d^4x \sqrt{|g|} \left[ \frac{M_P^2}{2} R + \frac{a}{2} R^2(1+ b \log (R/\mu^2))^{-1} \right] \, .  
\end{equation}
Here it was argued that the logarithmic terms arise from resuming the higher-order scalar curvature terms which are not included in the ansatz \eqref{eq:RGaction}. This particular form of the higher-derivative curvature terms has a clear motivation based on the structure of the NGFP underlying asymptotic safety where the logarithmic corrections arise naturally \cite{Dietz:2012ic,Demmel:2015oqa}. While the solution of the RG equation for the couplings $a$,$b$ connecting the effective dynamics at trans-Planckian and inflationary scales is currently not available, it is nevertheless conceivable that \eqref{refinedStarobinsky} constitutes a good approximation at $k_{\rm infl}$ as well. The fact that both \eqref{eq:RGaction} and \eqref{refinedStarobinsky} give rise to very interesting cosmological models which may be testable in future CMB experiments (the later predicting a tensor-to-scalar ratio in the range of $0.001 \le r \le 0.10$) make it tempting to combine the two approaches and verify that the structure \eqref{refinedStarobinsky}  indeed arises from a FRG running similar to the one constructed in this work.

The rest of the paper is organized as follows. In section~\ref{sec:observational_constraints} we derive the observational constraints reported in table~\ref{tab:constraints}. In section~\ref{sec:rg_structure} we discuss the structural properties of the RG phase diagram of the $R^2$-model. In section~\ref{sec:physical_rg_flows} we impose the observational constraints on the RG flow, and show that there indeed exists a RG trajectory satisfying all of them. The main body of the paper concludes with a discussion of the implications of these results. Technical details on converting $f(R)$-type Lagrangians to the Einstein frame and the $\beta$-functions underlying the analysis obtained in \cite{Machado:2007ea} have been relegated to appendix~\ref{sec:tensor_scalar_to_fr_link} and appendix~\ref{sec:rg_machinery}, respectively.

\section{Observational constraints on gravity}
\label{sec:observational_constraints}
The overall goal of this work is the construction of an RG trajectory passing through all points listed in table~\ref{tab:constraints}. In this section we start by deriving the values of the couplings at the corresponding energy scales based on cosmological observations made over different distance scales. 

At this stage the following introductory remark is in order. Table~\ref{tab:constraints} may suggest that only one parameter is constrained at each given energy scale. The derivation of these values assumes that the other parameters take ``reasonable values'' at the specified scale (the meaning of this will be made  precise below when discussing the individual constraints). In particular, it is assumed that Newton's coupling $G_k$ does not run significantly between inflationary scales and the laboratory scales where it is currently measured. Denoting the laboratory value of Newton's constant by $G$, we introduce the reduced Planck mass $M_{P}$ by the standard relation
\begin{equation}
		M_{P}	=	(8 \pi G)^{-1/2} = 2.4 \times 10^{27} \,\, \text{eV}
	\,\text{.}
	\label{eq:NewtonPlanck}
\end{equation}
At this stage we adopt these properties as a working hypothesis. Once a viable RG trajectory is found, we can check a posteriori that these working assumptions are indeed met.

\subsection{Observational constraints from primordial cosmology}\label{sec:primordialconstraints}
Assuming that the inflationary phase in the early universe originates from the $R^2$-term (Starobinsky inflation), the parameter $B$ can be constrained by early time cosmological observations. In the context of primordial cosmology, it is generally assumed that  the energy density of matter and of the cosmological constant are negligible, meaning that these components do not influence the background dynamics significantly. 
We take as a working assumption that the RG flow of the cosmological constant is such that it does not spoil this approximation and the contribution of $\Lambda_{k_{\text{infl}}}$ to the early universe dynamics can be considered as subdominant. Then, once a viable RG trajectory is selected, we can check a posteriori that this approximation is indeed met.
 
Neglecting the contribution of $\Lambda$, the action \eqref{eq:action} reduces to the so-called Starobinsky model \cite{Starobinsky:1980te, Mukhanov:1981xt, Starobinsky:1983zz} for inflation.
Inflationary models can be constrained by using observations of the cosmic microwave background, as done most recently by the Planck collaboration \cite{Ade:2015lrj}. Because these constraints rely on the inferred properties of primordial perturbations when they left the Hubble horizon, we take the  Hubble parameter at that time as the relevant energy scale:\footnote{What observations actually provide is an upper bound on the value of the Hubble parameter. In the Einstein frame, this bound is given by $H_{\text{infl}}<3.6 \times 10^{-5} M_{P}$, see ref.\ \cite{Ade:2015lrj}. The fact that this is an upper bound rather than an estimate is because $H_{\text{infl}}\propto \sqrt{\frac{r}{0.1}}$, and we only have upper bounds on the tensor-to-scalar ratio $r$. The order of magnitude estimate that we give in eq. \eqref{eq:kinfl} is based on the general expectation that $r$ should not be much smaller than its current upper bound. For Starobinsky inflation based on the model \eqref{eq:RGaction} the prediction for the tensor-to-scalar ratio is $r = 12/N^2 \approx 0.004$ which in the slow-roll approximation translates to $H_{\rm infl} = 7.2 \times 10^{-6} M_{P}$. In the Jordan frame the value for $H_{\text{infl}} = 5.5 \times 10^{-5} M_{P}$ turns out to be higher, see \cite{Netto:2015cba} for a detailed discussion.} 
\begin{equation}
		k_{\text{infl}}	=	H_{\text{infl}}	\simeq 10^{22} 
	\,\text{eV}
	\text{.}
	\label{eq:kinfl}
\end{equation}

Constraints on inflationary models  usually refer to parameters of the inflaton field potential  $V(\varphi)$ \cite{Ade:2015lrj, Martin:2013tda}. We can apply these results to the $R^{2}$-model we are interested in, since at the  classical level $f(R)$-gravity can be rewritten using the equations of motion into an action for gravity coupled to a scalar field $\varphi$, with  a potential $V(\varphi)$. 
Following the derivation of appendix~\ref{app:FRtoVphi}, the action \eqref{eq:action} with $\Lambda=0$ leads to a scalar potential
\begin{equation}
	V(\varphi)	=	\frac{M_{P}^{2}}{8B}   \left[ 1 - e^{-\sqrt{2/3}  \varphi/M_{P}}\right]^{2}  
	\,\text{,}
	\label{eq:scalarPotential}
\end{equation}
which characterizes the Higgs inflation model \cite{Bezrukov:2007ep, Barvinsky:2008ia}. Recent constraints on this model are reported in \cite{Starobinsky:1983zz,Netto:2015cba,Liu:2018aa}:\footnote{As explained in appendix~\ref{sec:tensor_scalar_to_fr_link}, we are using an action with opposite overall sign with respect to the one conventionally used in cosmology.}\footnote{We thank an anonymous referee and the authors of \cite{Martin:2013tda} for correspondence confirming the correctness of this value.}
\begin{equation}
		M\simeq 3.3 \times 10^{-3} M_{P}
	\,\text{,}
	\label{eq:PlanckConstraint}
\end{equation}
where  $M^{4}=-\frac{M_{P}\,^{2}}{8 B}$. When written in terms of the parameters \eqref{eq:RGaction} this constraint implies
\begin{equation}
		M_{P}^{2} \, B_k \simeq -1 \times 10^{9} 
		\,\text{,}
\end{equation}
with $k$ taken at horizon crossing, $k=k_{\text{infl}}$.

\subsection{Late time cosmology}\label{sec:latetimeconstraint}
In the previous subsection we showed that early time cosmology provides a constraint on the $R^{2}$ coupling at high energy scales, $k_{\text{infl}}$. In contrast, late time cosmology is sensitive to the value of the cosmological constant at very low energy scales corresponding roughly to the current value of the Hubble parameter. Thus measurements of the cosmological constant are done at an energy scale
\begin{equation}
		k_{\text{Hub}} = 10^{-33}
		\,\,\text{eV}
	\,\text{.}
\end{equation}

While the constraint on $B$ derived in the previous section relies on the assumption that the $R^{2}$ term dominates the universe dynamics at early times, here we make the complementary assumption that late-time dynamics is only sensitive to the standard Einstein-Hilbert term $R$  and the cosmological constant $\Lambda$ (i.e. we rely on the standard $\Lambda$CDM cosmological model). Because the curvature is very small, the $R^{2}$ term can be considered negligible, as long as the RG flow does not drive the coupling $B$ to extremely large values. 
 
Current late-time cosmological observations are fully compatible with a universe dynamics governed at late times by a $R-2\Lambda$ action \cite{Ade:2015xua}. Specifically, the cosmological constant density parameter takes the value
\begin{equation}
		\Omega_{\Lambda} \simeq 0.7
		\,\text{.}
\end{equation}
Together with the current value of the Hubble parameter \cite{Patrignani:2016xqp},
\begin{equation}
		H_{0}\simeq 70 \,\text{km} \,\text{s}^{-1} \,\text{Mpc}^{-1} 
	\,\text{,}
\end{equation}
this allows to estimate the cosmological constant as follows:\footnote{Note that the main uncertainty on the value of the cosmological constant comes from the tension in competing estimates of the Hubble constant, from the CMB  \cite{Ade:2015xua} and from  astrophysical observations (e.g. \cite{Riess:2016jrr}). However, this uncertainty is not significant for the purposes of our analysis.}
\begin{eqnarray}
		\Omega_{\Lambda} & \equiv& \frac{\rho_{\text{vac}}}{\rho_{c}} = \frac{\Lambda}{8\pi G}\frac{8\pi G}{3 H_{0}^{2}}
	\\\nonumber\\
		\Rightarrow\Lambda &=& \Omega_{\Lambda}\cdot 3H_{0}^{2} \simeq  4 \times 10^{-66} \,\,\text{eV}^{2} 
	\,\text{.}
\end{eqnarray}

\subsection{Newton's gravitational constant}

Current  estimates of Newton's gravitational constant are based on laboratory experiments, made on scales of about $10^{-2}-10^{0}$ m \cite{Patrignani:2016xqp}, corresponding to energies of $10^{-4}-10^{-6}$ eV.  We use as reference scale the intermediate value:
\begin{equation}
		k_{\text{lab}} \simeq 10^{-5} \text{ eV}
	\,\text{.}
\end{equation}
The most up-to-date value of the Newton coupling is provided in \cite{Patrignani:2016xqp}:
\begin{equation}
		G	=	6.7 \times 10^{-57} \text{ eV}^{-2}
	\text{.}
	\label{eq:2.12}
\end{equation}
Since this value is obtained from a local measurement where the laws of gravity are captured by Newtonian gravity, the result \eqref{eq:2.12} is not sensitive to the value of the cosmological constant or to the $R^{2}$ coupling.

\section{RG structure}
\label{sec:rg_structure}
The second ingredient in our analysis is the gravitational RG flow projected onto actions of the form \eqref{eq:RGaction}. This flow relates the values of the coupling constants at different energy scales $k$. Concretely, we base our analysis on the flow equation derived in \cite{Machado:2007ea} which is reviewed in appendix~\ref{sec:rg_machinery}. Our results extend earlier work \cite{Rechenberger:2012pm}.

The scale-dependence of the couplings $G_k, \Lambda_k$ and $B_k$ is most conveniently analyzed in terms of their dimensionless counterparts
\begin{equation}
		\lambda_k \coloneqq \Lambda_k \, k^{-2}
	\,\text{,}
	\quad
		g_k \coloneqq G_k \,k^{2}		
	\,\text{,}
	\quad
		b_k \coloneqq B_k \, k^{2}		
	\,\text{.}
	\label{eq:dimensionless_couplings}
\end{equation}
The $k$-dependence of these couplings is governed by their $\beta$-functions
\begin{equation}
		\partial_t g_k = \beta_g(g,\lambda,b) 
	\,\text{,}\qquad
		\partial_t \lambda_k = \beta_\lambda(g,\lambda,b)
	\,\text{,}\qquad
		\partial_t b_k = \beta_b(g,\lambda,b)
	\,\text{,}
	\label{eq:beta1a}
\end{equation}
where $t = \log (k/k_0) $ denotes the logarithmic RG time and $k_0$ is an arbitrary reference scale. The functions $\beta_i(g,\lambda,b)$ are obtained by solving the system \eqref{eq:beta2}. In order to get an idea about the RG trajectory realized by Nature we first investigate the fixed point and singularity structure of the $\beta$-functions in subsection~\ref{sub:fixed_points} before constructing sample trajectories in subsection~\ref{sub:trajectory_shapes_for_initial_conditions_in_relevant_quadrants_of_g_0_plane}. The main result of this section is given in figure~\ref{fig:singplot}.

\subsection{Fixed points, singularities and separation lines}
\label{sub:fixed_points}
By definition a renormalization group fixed point corresponds to a point $\{g_*\}$ in the space of coupling constants where all $\beta$-functions vanish simultaneously. Depending on whether this point corresponds to a free or interacting theory, one distinguishes between the so-called Gaussian fixed point (GFP) and non-Gaussian fixed points (NGFPs). Investigating the system \eqref{eq:beta2}, one finds that there are two fixed points relevant for the present analysis. The GFP is located at
\begin{equation}\label{GFP1}
	\lambda_* 	=	0	\,\text{,}
	\quad
	g_* 		=	0	\,\text{,}
	\quad
	b_* 		=	0	\,\text{.}
\end{equation}
In addition the system exhibits a NGFP situated at
\begin{equation}\label{NGFP1}
	\lambda_* 	=	0.133	\,\text{,}
	\quad
	g_* 		=	1.59	\,\text{,}
	\quad
	b_* 		=	0.119	\, \text{.}
\end{equation}
The dimensionless coupling multiplying the $R^2$ term then takes the value\footnote{The positive value of this coupling leads to ``stable inflation''. The transition to a negative coupling occurs along the RG flow which then realizes the ``unstable inflation''-scenario analyzed in \cite{Netto:2015cba}.}
\begin{equation}
\frac{b_*}{16 \pi g_*} = 1.5 \cdot 10^{-3} \, . 
\end{equation}

The flow in the vicinity of a RG fixed point can be studied by linearizing the $\beta$-functions around this point. The properties of the linearized flow are captured by the stability matrix ${\bf B}_{ij} := \left. \partial_{g_j} \beta_{g_i} \right|_{g = g_*}$. Defining the critical exponents $\theta_i$ as minus the eigenvalues of ${\bf B}_{ij}$, eigendirections whose $\theta_i$ come with a positive real part attract the RG flow for increasing $k$ while critical exponents with a negative real part are UV repulsive. Evaluating the stability matrix for the GFP yields the critical exponents
\begin{equation}
 \mbox{GFP:} \qquad \theta_1 = +2 \,\text{,} \quad \theta_2 = -2 \, , \quad \theta_3 = -2 \, . 
\end{equation}
Combining this information with the associated eigendirections shows that the GFP is UV-attractive in the $g=0$ plane and UV-repulsive in the two other eigendirections. The analogous analysis for the NGFP yields \cite{Machado:2007ea}
\begin{equation}
\mbox{NGFP:} \qquad \theta_{1,2}	=	1.26	\pm 	2.45 \imath 	\,,
\quad
\theta_3		=	27.0							\, ,
\end{equation}
indicating that the NGFP acts as an UV-attractor for RG trajectories entering its vicinity. The large positive eigenvalue $\theta_3$ is typical for the $R^2$ system (see e.g. \cite{Lauscher:2002sq}). Its value reduces significantly once higher-order curvature terms are included \cite{Codello:2007bd,Machado:2007ea,Falls:2014tra}. These results also show that the critical exponents associated with curvature terms of order $R^3$ and higher come with a negative real part, so that the present approximation most likely includes all the free parameters occurring in asymptotically safe gravity.

\begin{figure}[t!]
	\centering
	\begin{subfigure}{.49\textwidth}
		\includegraphics[width=.93\textwidth]{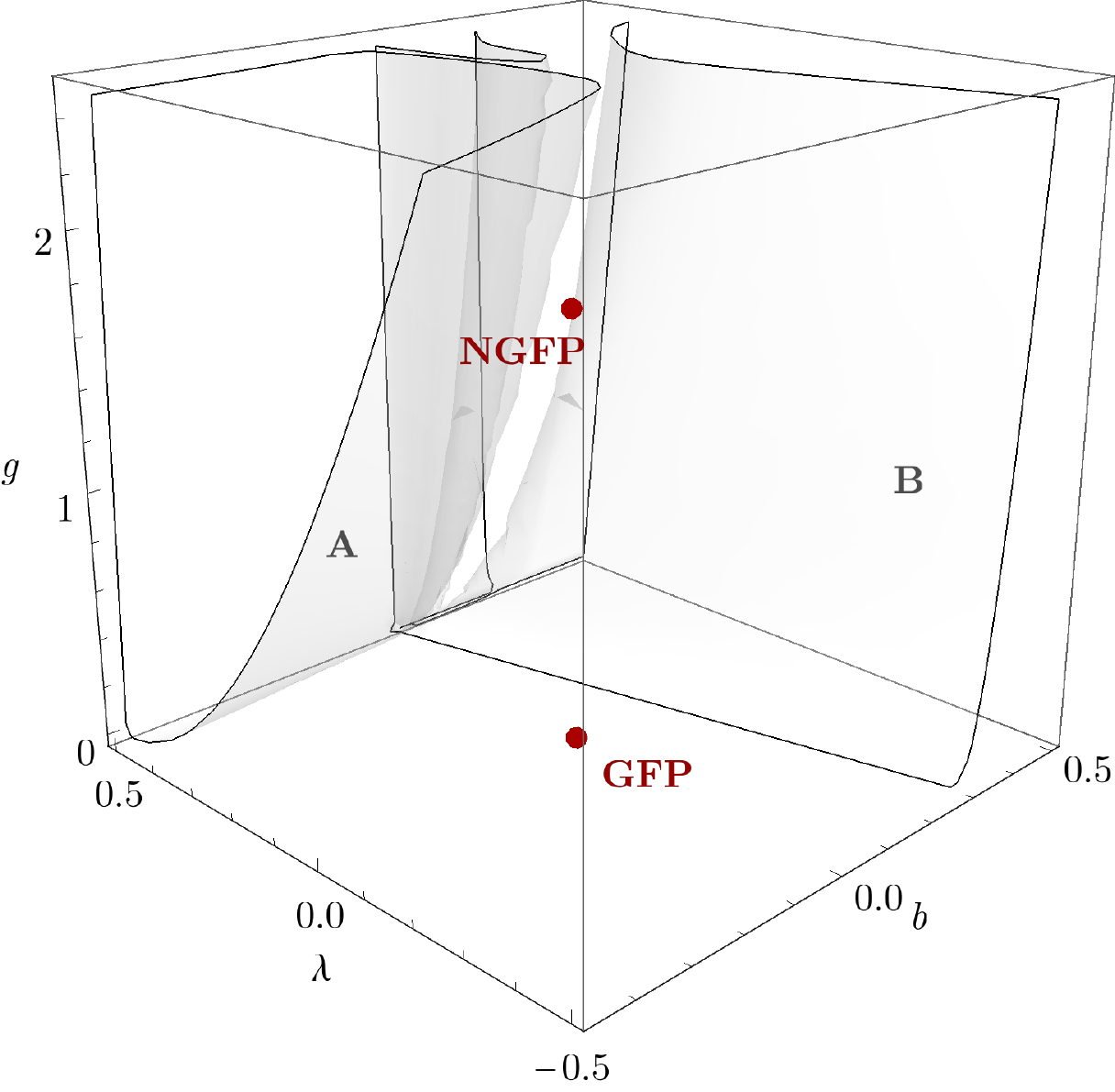}
		\caption{}\label{fig:singplot:a}
	\end{subfigure}
	\begin{subfigure}{.49\textwidth}
		\includegraphics[width=.93\textwidth]{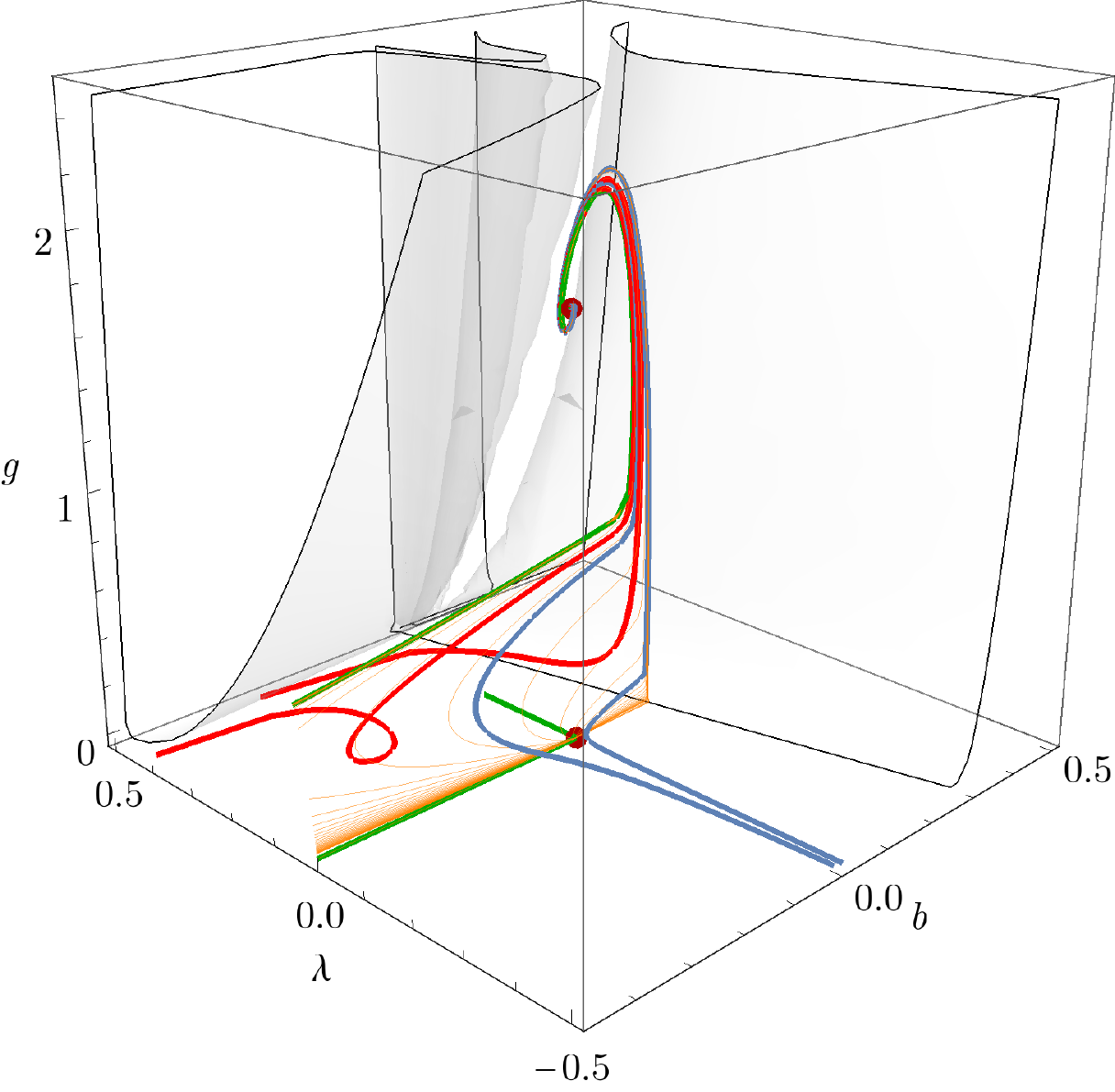}
		\caption{}\label{fig:singplot:b}
	\end{subfigure}
	
	\caption{
		Overview of the flow diagram.
		\ref{fig:singplot:a}: the Gaussian fixed point \eqref{GFP1} and non-Gaussian fixed point \eqref{NGFP1} are marked by the red dots. The singular planes stretch out in negative-$b$ direction (A) and in negative-$\lambda$ direction (B).
		\ref{fig:singplot:b}: selected RG trajectories.
		The red curves lead to a positive IR value of the cosmological constant and are denoted as trajectories of Type IIIa. These trajectories eventually terminate in the singular plane (A).
		The blue curves denote trajectories of Type Ia and possess a negative IR value of the cosmological constant. These curves avoid any singularities.
		The surface spanned by the orange curves marks the trajectories of Type IIa. Their IR limit is given by the GFP. These solutions separate the trajectories of Type Ia and IIIa.
		Finally, the green curve marks the trajectory that meets the observational constraints.
	}
	\label{fig:singplot}
\end{figure}

Besides its fixed points, the singular loci of the $\beta$-functions play an essential role in constructing the phase diagram. The region containing the GFP and the NGFP is bounded by two planes where $\beta_i(g,\lambda,b)$ is infinite. These planes are shown in the left panel of figure~\ref{fig:singplot} and labeled by the letters ``A'' and ``B'', respectively. The surface A is parabola-shaped and extends approximately parallel to the  $b$-axes. This surface constitutes a possible termination point for RG-trajectories flowing to positive values of $\lambda$ at small values $k$. The surface B bounds the region containing the GFP and NGFP towards positive values $b$. Typically RG trajectories do not terminate in this surface since
they are repelled once they come close to it. 

Finally, one finds that the $\beta$-function for the Newton coupling, $\beta_g(g,\lambda,b)$, vanishes as $g=0$. As a consequence RG trajectories can not cross the $g=0$ plane. Since the observed value of the Newton coupling is positive at laboratory scales, this feature limits the physically interesting sector of the phase diagram to positive values $g_k$.

\subsection{Construction of sample trajectories}
\label{sub:trajectory_shapes_for_initial_conditions_in_relevant_quadrants_of_g_0_plane}
Following up on determining the relevant fixed point and singularity structures linked to the RG flow of the $R^2$-action \eqref{eq:RGaction} we proceed by constructing explicit sample solutions by integrating the flow equations \eqref{eq:beta1a} numerically. Our primary focus is on RG trajectories which emanate from the NGFP \eqref{NGFP1} at large values of $t$ and undergo a crossover to the GFP as $t$ is lowered. Typical examples exhibiting this behavior are shown in the right panel of figure~\ref{fig:singplot}. Following the discussion \cite{Reuter:2001ag,Reuter:2004nx,Rechenberger:2012pm} it is this type of solutions which give rise to a classical regime resembling general relativity at low energy.

Following the classification in \cite{Reuter:2001ag} for the Einstein-Hilbert truncation, a useful discriminator for the solutions is the IR value of the cosmological constant. Similarly to the Einstein-Hilbert classifications one encounters trajectories which flow towards negative (Type~Ia, blue curves) and positive $\Lambda$ (Type IIIa, red curves) as $t$ is lowered. The former are typically well-defined for all values of $t$ while the latter terminate in the singular locus A. Finally, there are solutions (orange lines) which start at the NGFP and end at the GFP as $t \rightarrow -\infty$. These trajectories give rise to a vanishing IR value of the cosmological constant. They span a two-dimensional plane which separates the solutions of Type Ia and Type IIIa. 

It is also instructive to discuss the flow of the 
$R^2$-coupling $b_k$ along these trajectories. In all cases, $b$ flows to zero in the IR. 
Let us first consider the trajectories that flow towards $b \to 0$ from positive $b$. Starting in the IR ($g_k \ll 1$) and following the flow into the UV, one finds that after approaching the GFP, these trajectories make a sharp turn towards positive $b$. As the trajectories approach the singular plane B, they are eventually repelled in the direction of the NGFP. This fixed point then provides the UV completion of the trajectory as $t \rightarrow \infty$.

Trajectories that approach $b_k \to 0$ from negative $b$ show a different behavior. Tracing the flow from IR to the UV, they first flow towards the GFP before making a turn towards negative $b$.
After obtaining a minimum value of $b$, they take a sharp turn and flow back into the direction of the GFP before entering the NGFP regime.
In contrast to the trajectories with positive $b$, the flow of this type of trajectories may be bounded by the singular plane A. This ceiling may prevent the solutions from reaching the basin of attraction of the NGFP so that they terminate at a finite value $t$ once they leave the $g \ll 1$ region.

\section{A complete cosmic history from Asymptotic Safety}
\label{sec:physical_rg_flows}
Based on the phase space analysis of the previous section, we are ready to check if there is an RG trajectory satisfying all the conditions listed in table~\ref{tab:constraints}. The values of the couplings listed in this table refer to different energy scales which turns the search of the corresponding RG trajectory into a rather complicated boundary value problem. In order to simplify the analysis we first convert this setup into an initial value problem in subsection~\ref{sect.4.1} before constructing the corresponding RG trajectory numerically in subsection~\ref{sect.4.2}. The main result of this section is the trajectory displayed in figure~\ref{fig:physicaltrajectory} which meets all cosmological requirements. 

\subsection{Initial values for the RG trajectory realized by Nature}
\label{sect.4.1} 
Our first task is to determine the value of $b_k$ and $\lambda_k$ at the scale $k_0 \equiv k_{\rm lab}$ which will be used to set up the initial value problem. Since inflation and the measurement of the cosmological constant occur well below the Planck scale, $g_k \ll 1$ in this regime. Thus an expansion of the $\beta$-functions in powers of $g_k$ which retains the leading quantum corrections is sufficient to capture all relevant features of the RG flow in this regime.\footnote{Essentially, this setup corresponds to describing the RG flow by a one-loop approximation of the $\beta$-functions. The results displayed in table~\ref{tab:physicaltrajectory} confirm that this is indeed a valid approximation in the regime of interest.} From a practical viewpoint, it is convenient to rewrite the $\beta$-functions \eqref{eq:beta1a} in terms of the new couplings
\begin{equation}
	\label{couplingredef}
		g_k \, , \quad u^1_k \equiv g_k \lambda_k \, , \quad u^2_k \equiv \frac{b_k}{16 \pi g_k} \, , 
\end{equation}
since this allows to find analytic expressions relating the values of the couplings at different scales $k$.

 We start with the expansion of the $\beta_g(g,u^1,u^2)$. This gives up to second order in $g$
\begin{equation}
		\beta_g(g,u^1,u^2)	=	2 g_k	-	 \frac{23}{24\pi}	g_k^2	+	\mathcal{O}(g_k^3)
	\text{,}
\end{equation}
which is independent of $u^1$ and $u^2$. The flow of $g_k$ can then be integrated analytically, yielding
\begin{equation}
g_k	=	\frac{48 \pi k^2 g_{k_0} }{48 \pi k_0^2	+ 23 g_{k_0} ( k^2 - k_0^2)}
\,\text{.}
\label{eq:gexpanded}
\end{equation}
We see that indeed, if $k \simeq k_0$, the dimensionful Newton's coupling is constant, up to corrections of order in $(k^2 - k_0^2)$. Thus it is convenient to identify $k_0 = k_{\rm lab}$ since this corresponds to the scale where $G_k$ is measured. Converting the measured value of the dimensionful Newton's constant into dimensionless quantities gives 
\begin{equation}\label{ginit}
g_{k_0}	=	6.71 \times 10^{-67}
\text{.}
\end{equation}

The expansion of $\beta_{u^1}(g,u^1,u^2)$ yields
\begin{equation}
\beta_{u^1}(g,u^1,u^2)
=
-	\frac{23}{12 \pi}	u^1_k \,	g_k
-	\frac{5}{12 \pi}	g_k^2
+	\mathcal{O}(g_k^3)
\text{.}
\end{equation}
Substituting the explicit solution for $g_k$, eq.\ \eqref{eq:gexpanded}, one finds that the scale-dependence of $u^1_k$ is given by
\begin{equation}
u^1_k
=
\frac{48 \pi \left(5 g_{k_0}^2 (1 - k^4/k_0^4) + 48 \pi u^1_{k_0} \right)}{\left(48 \pi + 23 g_{k_0} (k^2/k_0^2 - 1)\right)^2}
\text{.}
\end{equation}
Combining the solutions of $g_k$ and $u^1_k$, allows to map the value of the cosmological constant measured at the Hubble scale $k_{\text{Hub}}$ to the value of $u^1_k$ at the laboratory scale
\begin{equation}\label{u1val}
u^1_{k_0} 
=
2.68 \times 10^{-122}
\text{.}
\end{equation}

Finally, the expansion of $\beta_{u^2}(g,u^1,u^2)$ yields
\begin{equation}
			\beta_{u^2}(g,u^1,u^2)
	=
			\frac{109}{2160 \pi^2}	+ \mathcal{O}(g_k)
	\text{.}
\end{equation}
The solution of this equation gives the typical logarithmic running of a marginal coupling at one-loop level
\begin{equation}
			u^2_k
	=
			\frac{109}{2160 \pi^2}	\log\left(\frac{k}{k_0}\right)	+ u^2_{k_0}
	\text{.}
\end{equation}
Inserting the measured values at $k_{\text{infl}}$ gives the initial value
\begin{equation}\label{u2val}
			u^2_{k_0}
	=
		-	5.0	\times 10^{8}
	\text{.}
\end{equation}
Combining the results \eqref{ginit}, \eqref{u1val}, and \eqref{u2val} with the definition \eqref{couplingredef} one readily arrives at the initial values for $g_k, \lambda_k$, and $b_k$
\begin{equation}\label{initialfinal}
g_{k_0} = 6.71 \times 10^{-67} \, , \quad 
\lambda_{k_0} = 3.99 \times 10^{-56} \, , \quad 
b_{k_0} = -1.7 \times 10^{-56}\, \text{.}
\end{equation}
These values serve as the starting point for integrating the flow equation numerically.
\subsection{The RG trajectory realized by Nature}
\label{sect.4.2}
In order to obtain the RG trajectory resulting from the initial conditions \eqref{initialfinal}, we now integrate the full $\beta$-functions numerically. 
For this, we  use the NDSolve routine in Mathematica. Since the initial starting point is very close to zero, we  increase the working precision to 124 digits. When the integration reaches a regime where the parameters take larger values, we decrease the precision to 25 digits. This allows us to track the RG flow from the classical regime up to the NGFP regime.

\begin{figure}
	\centering
	\includegraphics[width=.7\textwidth]{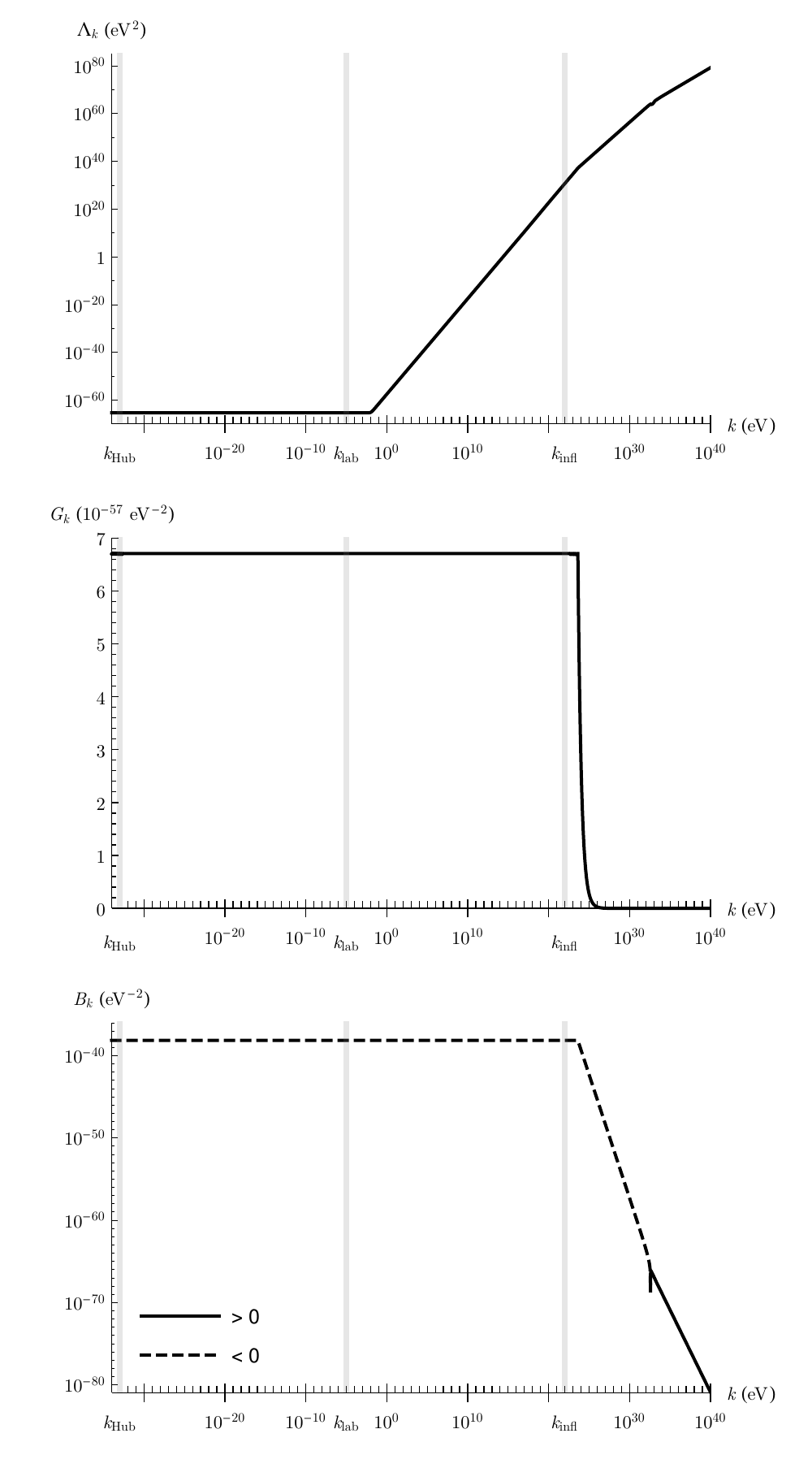}
	\caption{RG trajectory passing through all points identified in table~\ref{tab:constraints}. Top panel: cosmological constant $\Lambda_k$. Middle panel: Newton's coupling $G_k$. Bottom panel: $R^2$ coupling $B_k$. The gray bars indicate the energy scales at which constraints are imposed.}
	\label{fig:physicaltrajectory}
\end{figure}

The resulting RG trajectory is depicted by the green curve in figure~\ref{fig:singplot:b}. The scale-dependence of the corresponding dimensionful couplings $\Lambda_k$, $G_k$, and $B_k$ is summarized in figure~\ref{fig:physicaltrajectory}. In table~\ref{tab:physicaltrajectory}, we  summarize the values of the couplings at  different relevant energy scales.
The existence of this RG trajectory constitutes the main result of this work.

\begin{table}
	\renewcommand{\arraystretch}{1.5}
	\centering
	\begin{tabular}{cccc}
		Energy scale					&	$\Lambda$	&	$G$ 			&	$B$
	\\
		(eV)								&	(eV$^2$)	&	($10^{-57}$ eV$^{-2}$)		&	($10^{-46}$ eV$^{-2}$)
	\\
		\hline
		$k_{\text{Hub}} \simeq 10^{-33}$	&	\boldmath{$4 \times 10^{-66}$}	&	$6.71$			&	$-1.7$
	\\
		$k_{\text{lab}} \simeq 10^{-5}$		&	$4 \times 10^{-66}$		&	\boldmath{$6.71$}	&	$-1.7$
	\\
		$k_{\text{infl}} \simeq 10^{22}$	&	$4 \times 10^{30}$		&	$6.71$			&	\boldmath{$-1.7$}
	\\
	\hline
	\end{tabular}
	\caption{Selected values of the RG flow satisfying the observational constraints. Within experimental uncertainty, the parameters are constant, except for the cosmological constant $\Lambda$. The bold values are constrained by observations, whereas the other values are results from the RG calculation.}
	\label{tab:physicaltrajectory}
\end{table}

The solution shown in figure~\ref{fig:physicaltrajectory} exhibits several remarkable features. First of all, we see that indeed the RG flow meets all observational constraints. This means that the expansion in small $g$ is valid, and that quantum effects  only play a role at energy scales beyond inflation. In fact, both the Newton's coupling and the $R^2$ coupling start to run at $10^{24}$ eV, which is  beyond the upper bound on the inflation scale, $10^{22}$ eV. Interestingly, the Newton's coupling reaches a very small value at the Planck scale $10^{27}$ eV. \footnote{Cosmological consequences of a Planck-scale-vanishing Newton's coupling were discussed from a different perspective  in \cite{Amelino-Camelia:2015dqa, Brighenti:2016xng}. We defer to a future work to investigate the relation between the two approaches.}

Secondly, we observe that $\Lambda_k$ starts to run at $10^{-2}$ eV, corresponding to a length scale of $10^{-4}$ m. At lower energy scales, $\Lambda_k$ is constant and equal to the value quoted in table~\ref{tab:constraints1}, thus not spoiling the agreement with current measurements, that take place over cosmological scales.

We close our discussion by verifying that our RG trajectory meets all the working assumptions made in section~\ref{sec:observational_constraints}. At the inflation scale, one has to check that $\Lambda_{k_{\text{infl}}} = 10^{30}\,\text{eV}^{2}$ does not affect the dynamics in this regime so that the constraint obtained on $B$ is valid. 
This amounts to the condition $\Lambda_k /	(B_k R^2) \ll 1$ 
evaluated at $k = k_{\text{infl}}$. 
Using that in an approximately de Sitter background the curvature $R$ is related to the Hubble parameter as $R\simeq12 H^{2}$ and the relation that $H = 	k_{\text{infl}}$, we find
\begin{equation}\label{infcheck}
			\Lambda_k / (B_k R^2)	\simeq	4 \times 10^{-15} \ll 1
	\text{.}
\end{equation}
Thus our working assumption is indeed valid in this regime.

The analysis of the late time cosmological evolution assumed that the $R^{2}$ term does not significantly affect the background universe evolution at $k_{\text{Hub}}$. Again making use of the fact that at late times we are in a quasi-de Sitter background so that $R \simeq 12 (k_{\text{Hub}})^2$, we find
\begin{equation}
			B_k R^{2}	\simeq 10^{-161} \;\text{eV}^{2} \, , \qquad
			\Lambda_k	\simeq	10^{-66} \;\text{eV}^{2} \, , \qquad
			R	\simeq	10^{-65} \;\text{eV}^{2}
	\text{.}
\end{equation}
Together with the result \eqref{infcheck} this establishes that our RG trajectory indeed fulfills all the working assumptions made when deriving table~\ref{tab:constraints}.

\section{Conclusion}
\label{sec:conclusion}
In this work we have studied the compatibility of the asymptotic safety mechanism with the laws of gravity observed at sub-Planckian scales. The main focus has been on obtaining a viable cosmological evolution. This question has already been subject to a significant amount of work. Starting from the seminal works \cite{Bonanno:2001hi,Bonanno:2001xi}, the cosmological evolution and its signatures have been studied using a modified dynamics resulting from scale-dependent couplings \cite{Guberina:2002wt,Babic:2004ev,Reuter:2005kb,Bonanno:2005mt,Bonanno:2007wg,Weinberg:2009wa,Bonanno:2010bt,Contillo:2011ag,Cai:2011kd}, dilaton-gravity \cite{Reuter:2003ca,Tronconi:2017wps}, Higgs-inflation inspired models \cite{Cai:2013caa,Saltas:2015vsc}, non-Gaussian fixed point driven inflation \cite{Bonanno:2012jy,Bonanno:2015fga,Bonanno:2018gck}, and anisotropic models \cite{DOdorico:2015jtl}.
All these investigations rely on a so-called renormalization group (RG) improvement which relates the RG scale $k$ to a physical quantity like the Hubble scale or the Ricci scalar $R$ in order to capture ``the leading quantum gravity effects'' of the system. These studies have raised the expectation that the very early universe undergoes a phase of power-law inflation and an almost flat scalar power spectrum \cite{Bonanno:2017pkg}.

In this work, we perform the first cosmological analysis within asymptotically safe gravity based on a ``first principles'' analysis. The gravitational dynamics is obtained by solving the flow equation for the effective average action at the $R^2$-level,  giving the explicit energy dependence of the three gravitational couplings contained in \eqref{eq:RGaction}. Thus the gravitational dynamics is obtained \emph{without invoking a scale-identification procedure}. Analyzing the space of solutions compatible with Asymptotic Safety gave rise to two striking insights. First, there are solutions which are compatible with the  values of the Newton coupling and cosmological constant observed, respectively, at laboratory and cosmological scales. Moreover, these constraints from late-time cosmology are compatible with an $R^2$-term which gives rise to a phenomenologically viable phase of inflation, without spoiling the late-time cosmological evolution. The seeds for this salient physics below the Planck scale are laid by the NGFP governing the gravitational dynamics at trans-Planckian scales: mapping out the surface of asymptotically safe RG trajectories establishes that there are solutions which besides being compatible with observed low-energy physics also realize an inflationary phase based on the resulting gravitational dynamics. Thus Asymptotic Safety may naturally give rise to Starobinsky inflation and there is no need to introduce an ad hoc inflaton field. In this sense, the construction is very economic in attributing highly desirable features in the early and late-time dynamics of the universe to quantum gravity effects.\footnote{For a similar discussion at the level of effective field theory see \cite{Codello:2015pga,Codello:2016neo}.}  

As a by-product our analysis constructed the realistic RG trajectory displayed in figure~\ref{fig:physicaltrajectory}.\footnote{For an earlier analysis including the observed values for $G$ and $\Lambda$, see \cite{Reuter:2004nx}.}
An intriguing feature of this solution is that the cosmological constant $\Lambda_k$ acquires an energy dependence already at $10^{-2}$ eV. This scale is well below the Planck scale where the energy dependence of the Newton coupling and the $R^2$-coupling sets in and already occurs in the vicinity of the Gaussian fixed point, i.e., way below the energy scale where the Newton coupling and the $R^2$-coupling exhibit an appreciable scale-dependence. As a consequence, one expects that fluctuations with a length scale of $10^{-4}$ m (or smaller) feel a different value of $\Lambda_k$. Assuming that a fluctuation of this size is generated at the same time as the cosmic microwave background, the subsequent cosmological evolution would expand such a structure to a size of approximately $1$ m, a scale that is well below the resolution of current large-scale structure surveys. On the other hand, currently observed scales in the cosmic microwave background power spectrum would have a wavelength of $10^{-4}$ m  during the inflationary phase, after horizon exit.  It would be interesting to investigate if this effect leaves an imprint on the structure formation in the early universe.

We close our discussion with the following remark. It is clear that (free) matter fields give rise to additional contributions in the $\beta$-functions \eqref{eq:beta1} \cite{Dona:2013qba,Christiansen:2017cxa,Alkofer:2018fxj}. Naturally, the properties of the non-Gaussian gravity-matter fixed points, including its position and stability coefficients, differ from the ones encountered in pure gravity. At least for the class of gravity-dominated gravity-matter fixed points these properties are, by definition, qualitatively similar to the ones of the pure gravity fixed point underlying this work. We thus expect that the scenario developed in this work essentially carries over to these cases as well. Naturally, it would be very interesting to study the scale-dependence of the cosmological constant taking contributions from phase transitions in the matter sector (like the electroweak symmetry breaking) into account, see e.g., \cite{Martin:2012bt,Sola:2013gha,Padilla:2015aaa} for an elaborate discussion. 
Since the resulting toy models potentially incorporating such effect become complex rather quickly \cite{Gies:2014xha,Eichhorn:2015kea,Eichhorn:2017als}, this is left for future work.


\bigskip
\acknowledgments
 F.S.\ thanks A.\ Bonanno and A.\ Platania for many fruitful discussions. The work by F.S.\ and C.R.\ is supported by the Netherlands Organization for Scientific Research (NWO) within the Foundation for Fundamental Research on Matter (FOM) grant 13VP12.

\appendix
\section{\texorpdfstring{$f(R)$-gravity in the Jordan and Einstein frame}{f(R)-gravity in the Jordan and Einstein frame}}
\label{sec:tensor_scalar_to_fr_link}
\label{app:FRtoVphi}
When analyzing the dynamical consequences of an $f(R)$-type gravity model, it is convenient to recast the theory into the Einstein frame. The resulting action then consists of an Einstein-Hilbert term supplemented by an additional scalar degree of freedom.\footnote{At the one-loop level the on-shell equivalence of the two formulations has recently been demonstrated in \cite{Ruf:2017xon,Ohta:2017trn}.} In this appendix we exhibit the relation between these two formulations, following \cite{Chiba:2003ir}.

We start from a generic (Lorentzian) $f(R)$-theory supplemented by a generic action for matter fields $S_{m}[g_{\mu\nu}]$
\begin{equation}
	S=\frac{M_{P}^{2}}{2}\int d^{4}x\sqrt{-g} f(R) +S_{m}(g_{\mu\nu})\,. \label{eq:F(R)}
\end{equation}
We keep track of the matter action, even though this part does not influence the dynamics during inflation and becomes relevant at later stages of the cosmological evolution. Subsequently, we introduce an the auxiliary field $\phi$ and rewrite the action \eqref{eq:F(R)} according to
\begin{equation}\label{eq1}
S=\frac{M_{P}^{2}}{2}\int d^{4}x\sqrt{-g} \,  \left[f(\phi)+f'(\phi)(R-\phi)\right] +S_{m}(g_{\mu\nu})\,,
\end{equation}
where the prime denotes a derivative with respect to the argument. Presupposing that $f''(\phi)\neq 0$, the field equations derived from
\eqref{eq1} include $\phi=R$, entailing that eqs.\ \eqref{eq:F(R)} and \eqref{eq1} are dynamically equivalent on-shell. If $f'(\phi) > 0$,  the  conformal transformation $f'(\phi)g_{\mu\nu}\equiv \tilde g_{\mu\nu}$ brings the gravitational part of the action into
 Einstein-Hilbert form
\begin{eqnarray} \nonumber
	S&=&\tfrac{M_{P}^{2}}{2}\int d^{4}x\sqrt{-\tilde g} \left[\tilde R-\frac{3}{2f'(\phi)^{2}}\tilde g^{\mu\nu}\tilde\nabla_{\mu}f'(\phi)\tilde\nabla_{\nu}f'(\phi) -\frac{1}{f'(\phi)^{2}}\left(\phi f'(\phi)-f(\phi)\right)\right]
	\\
	&& \quad \label{eq2}
	 +S_{m}(\tilde g_{\mu\nu}/f'(\phi))\,.
\end{eqnarray}
The kinetic term for the scalar field $\phi$ can be brought into canonical form by introducing the new field $\varphi$ according to
\begin{equation}
	f'(\phi)\equiv \eulere^{\sqrt{2/3} \varphi/M_{P}}\, . \label{eq:varphi}
\end{equation}
If $f'(\phi)$ is monotonic, this relation implicitly defines a map $\phi(\varphi)$. Substituting this map into the action \eqref{eq2} 
yields
\begin{equation}
	S=\int d^{4}x\sqrt{-\tilde g} \left[\frac{M_{P}^{2}}{2} \tilde R-\frac{1}{2}(\tilde\nabla\varphi)^{2}-V(\varphi)\right]  +S_{m}(\tilde g_{\mu\nu}/F')\,,
\end{equation}
with the scalar potential given by
\begin{equation}
	V(\varphi)= \frac{M_{P}^{2}}{2f'(\phi(\varphi))^{2}}\Big(\phi(\varphi) f'(\phi(\varphi))-f(\phi(\varphi))\Big)\,.
\end{equation}

For the $f(R)$-type action \eqref{eq:action}, the transformation from Jordan to Einstein frame can be carried out explicitly. For this specific case\footnote{Note that we are working with an action that differs from  \eqref{eq:action} by an overall minus sign. To make up for this, in the main text in \eqref{eq:scalarPotential} we write the potential $V(\varphi)$ with opposite sign with respect to the one we get here.}
\begin{equation}\label{fRans}
	f(R)=-2 \Lambda+ R-B R^{2}\,.
\end{equation}
For $B=0$ the theory already is in the Einstein frame, so we assume that $B \not = 0$. Evaluating \eqref{eq:varphi} for $f(\phi)$ which is quadratic in $\phi$ leads to the following relation between the scalar fields:
\begin{equation}
	1- 2 B \phi = e^{\sqrt{2/3} \varphi/M_{p}} \; \; \; \Rightarrow \; \; \; \phi = \frac{1-e^{\sqrt{2/3} \varphi/M_{p}}}{ 2 B}\, . 
\end{equation}
Finally, the scalar potential resulting from \eqref{fRans} is
\begin{eqnarray}
	V(\varphi)&=&\frac{M_{P}^{2}}{2}   \left[- \frac{1}{4 B} +\frac{1}{2 B} e^{-\sqrt{2/3}  \varphi/M_{P}}  +  \left(-\frac{1}{4B}+2\Lambda\right) e^{-2\sqrt{2/3}  \varphi/M_{P}}\right] 
	 \,.\label{eq:scalarPotentialLambda}
\end{eqnarray}
This potential provides the starting point for analyzing the cosmological dynamics in section~\ref{sec:observational_constraints}.

\section{RG machinery}
\label{sec:rg_machinery}
The most important tool for constructing approximate solutions of the renormalization group flow of gravity is the functional renormalization group equation (FRGE) for the effective average action (EAA) $\Gamma_k$ \cite{Wetterich:1992yh,Morris:1993qb,Reuter:1993kw,Reuter:1996cp}. Schematically, the FRGE takes the form
\begin{equation}
			\partial_t 	\Gamma_k 
	=
			\frac{1}{2}	\text{Str}	\left[	\left(	\Gamma_k^{(2)}	+	\mathcal{R}_k 	\right)^{-1}	\partial_t\mathcal{R}_k	\right]
	\,,
	\label{eq:FRGE}
\end{equation}
where $\Gamma_k^{(2)}$ is the Hessian of the EAA and $t \coloneqq \ln(k/k_0)$ is the RG time. The regulator $\mathcal{R}_k$ suppresses field modes with momentum $p^2 \lesssim k^2$, while it vanishes for $p^2 \gtrsim k^2$. The supertrace denoted by Str sums over field components while providing necessary minus signs for ghost field contributions. 
The present work focuses on approximations where the gravitational part of $\Gamma_k$ is given by a $f(R)$-type gravitational action. Flow equations governing the scale-dependence of this function have been derived in \cite{Codello:2007bd,Machado:2007ea,Benedetti:2012dx,Falls:2014tra,Ohta:2015efa} and their properties have been analyzed in detail by various groups (see \cite{Benedetti:2012dx,Dietz:2012ic,Dietz:2013sba,Demmel:2014hla,Demmel:2015oqa,Ohta:2015fcu,deBrito:2018jxt} for selected works and further references). 

The present work builds on the flow equation derived in \cite{Machado:2007ea}. In order to obtain a self-contained manuscript, the main features and results of this construction are reviewed in this appendix.

In the remainder of this section, we discuss the particular setup that is used in \cite{Machado:2007ea}. The construction of the FRGE for gravity heavily relies on the background field formalism. The quantum metric $g_{\mu \nu}$ is decomposed into a fixed background metric $\bar g_{\mu \nu}$ and a fluctuating metric $h_{\mu \nu}$ using a linear split:
\begin{equation}
			g_{\mu \nu}
	=
			\bar g_{\mu \nu}	+	h_{\mu \nu}
	\,.
	\label{eq:linear_split}
\end{equation}
The full ansatz for the EAA is then taken to be
\begin{equation}
			\Gamma_k [g ; \bar g]
	=
			\bar \Gamma_k [g]	+	S_{\text{gf}}[g - \bar g ; \bar g]	+	S_{\text{gh}}	+	S_{\text{aux}}
	\,,
	\label{eq:ansatz}
\end{equation}
with
\begin{equation}
			\bar \Gamma_k[g]
	=
			\frac{1}{16	\pi G_k}	\int d^4 x 	\sqrt{g}	f_k(R)
	\,.
	\label{eq:f(r)_ansatz}
\end{equation}
This term is supplemented by a scale-independent gauge-fixing term implementing geometrical gauge in the Landau limit. The ansatz for $\Gamma_k$ is then completed by the corresponding ghost action and auxiliary fields exponentiating Jacobians arising from field redefinitions. Details on these terms can be found in the original article \cite{Machado:2007ea}.

Substituting the ansatz \eqref{eq:ansatz} into the FRGE and projecting the resulting flow on functions of the scalar curvature results in a partial differential equation governing the scale-dependence of $f_k(R)$. This
equation is most conveniently written in terms of the dimensionless quantities
\begin{equation}
	r := k^{-2}	R 	\quad	\text{and}	\quad	\mathcal{F}_k(r)	:=	\frac{1}{16	\pi G_k}	\, k^{-4} \,	f_k(r k^2) \, . 
	\label{eq:dimless_quantities}
\end{equation}
The partial differential equation satisfied by $\mathcal{F}_k$ then reads 
\cite{Machado:2007ea} 
\begin{equation}
			384	\pi^2	\left(	\partial_t 	\mathcal{F}_k 	+	4	\mathcal{F}_k - 2 r \mathcal{F}_k^{\prime}	\right) 
	=
			\sum_{i=1}^6 c_i
	\label{eq:flow_equation}
\end{equation}
with the $c_i$ given by
\begin{align}
			c_1 
	&=
			\Big[5 r^2 \theta \left ( 1 - \tfrac{r}{3} \right ) 
		-	(12 + 4 r - \tfrac{61}{90} ) r^2 \Big ]	\Big[1 - \tfrac{r}{3} \Big ]^{-1}
	\,,
	\\				
			c_2 
	&=
			10 r^2 \theta \left ( 1 - \tfrac{r}{3} \right )
	\,,	
	\\
			c_3 
	&= 	
			\Big[10 r^2 \theta \left ( 1 - \tfrac{r}{4} \right )
		-	r^2 \theta \left ( 1 + \tfrac{r}{4} \right ) 
		-	(36 + 6 r - \tfrac{67}{60} r^2 ) \Big]	\Big[ 1 - \tfrac{r}{4} \Big ]^{-1} 
	\,, 
	\\
			c_4 
	&= 	
			\Big[ \eta_f \left ( 10 - 5 r - \tfrac{271}{36} r^2 + \tfrac{7249}{4536} r^3 \right ) 
		+ 	(60 - 20 r - \tfrac{271}{18} r^2 ) \Big]	\Big[ 1 + \tfrac{\mathcal{F}_k}{\mathcal{F}_k^{\prime}} - \tfrac{r}{3} \Big]^{-1} 
	\,, 
	\\ \nonumber
			c_5 
	&=	
			\tfrac{5 r^2}{2} \Big[ \eta_f \left( \left( 1 + \tfrac{r}{3} \right ) \theta \left( 1 + \tfrac{r}{3} \right)
		+	\left( 2 + \tfrac{r}{3} \right) \theta \left( 1 + \tfrac{r}{6} \right) \right) 
		+ 	2 \theta \left( 1 + \tfrac{r}{3} \right) + 4 \theta \left( 1 + \tfrac{r}{6} \right) \Big] \times \\ & \qquad \quad \times
			\Big[ 1 + \tfrac{\mathcal{F}_k}{\mathcal{F}_k^{\prime}} - \tfrac{r}{3} \Big]^{-1} 
	\,, 
	\\ \nonumber
			c_6 
	&= 	
			\Big[ \mathcal{F}_k^{\prime} \, \eta_f \, \left( 6 + 3 r + \tfrac{29}{60} r^2 + \tfrac{37}{1512} r^3 \right) 
		+ 	(\partial_t \mathcal{F}_k^{\prime\prime} - 2 r \mathcal{F}_k^{\prime\prime\prime}) \left(27 - \tfrac{91}{20} r^2 - \tfrac{29}{30} r^3
		- 	\tfrac{181}{3360} r^4 \right) 
	\\ \nonumber
		& \quad +	\mathcal{F}_k^{\prime\prime} \left(216 - \tfrac{91}{5} r^2 - \tfrac{29}{15} r^3 \right) 
		+	\mathcal{F}_k^{\prime} \left(36 + 12 r + \tfrac{29}{30} r^2 \right) \Big] \times \\ & \qquad
			\times \Big[2 \mathcal{F}_k + 3 \mathcal{F}_k^{\prime} \left(1 - \tfrac{2}{3} r \right) + 9 \mathcal{F}_k^{\prime\prime} 
			\left(1 - \tfrac{r}{3} \right)^2 \Big]^{-1} 
	\,.	
\end{align}
Here, a prime denotes a derivative w.r.t. the dimensionless curvature scalar $r$ and $\eta_f$ is the anomalous dimension of $f^{\prime}_k(R)$, 
\begin{equation}
			\eta_f
	=
			\frac{1}{\mathcal{F}^{\prime}_k}	(
			\partial_t 	\mathcal{F}_k^{\prime}
		+	2	\mathcal{F}^{\prime}_k
		-	2	r 	\mathcal{F}^{\prime\prime}_k	) \, . 
	\label{eq:anomalous_dimension}
\end{equation}

Using the dimensionless variables introduced in eq.\ \eqref{eq:dimensionless_couplings}, the dimensionless function $\mathcal{F}_k(r)$ corresponding to the action \eqref{eq:RGaction} is 
\begin{equation}
			\mathcal{F}_k(r)
	=
			\frac{1}{16 \pi g_k}	(
			2	\lambda_k
		-	r
		+	b_k 	r^2	)
	\,.
	\label{eq:ansatz_phi}
\end{equation}
Substituting this expression into \eqref{eq:flow_equation} and subsequently expanding the result in a power series around $r=0$ the $\beta$-functions for the couplings $\lambda_k$, $g_k$ and $b_k$ can be read off from the three lowest order terms in this expansion. Concretely,
\begin{equation}
\partial_t g_k = \beta_g(g,\lambda,b) \, , \qquad
\partial_t \lambda_k = \beta_\lambda(g,\lambda,b) \, , \qquad
\partial_t b_k = \beta_b(g,\lambda,b) \, ,
\label{eq:beta1}
\end{equation} 
where the $\beta_i$ are obtained as the solution of the following linear system of equations
\begin{subequations}
\begin{multline}
		-	\frac{6 \left(9 g_k \beta_b-9 b_k \beta_g+72 b_k g_k+\beta_g-8 g_k\right)}{g_k (18 b_k+4 \lambda_k-3)}
	\\
		+	\frac{48 \left(-\pi  \lambda_k \beta_g+\pi  g_k \beta_\lambda+4 \pi  g_k \lambda_k\right)}{g_k^2}
		-	\frac{80-\frac{10 \beta_g}{g_k}}{1-2 \lambda_k}+48
		=	0
	\,\,\text{,}
\end{multline}
\begin{multline}
		-	\frac{-20 \beta_b+40 b_k+\frac{5 \beta_g}{g_k}-30}{1-2 \lambda_k}
		-	\frac{3 \left(4 g_k \beta_b-4 b_k \beta_g+24 b_k g_k+\beta_g-6 g_k\right)}{g_k (18 b_k+4 \lambda_k-3)}
	\\
		-	\frac{36 b_k \left(9 g_k \beta_b-9 b_k \beta_g+72 b_k g_k+\beta_g-8 g_k\right)}{g_k (18 b_k+4 \lambda_k-3)^2}
		-	\frac{2 (6 b_k \lambda_k-1) \left(80-\frac{10 \beta_g}{g_k}\right)}{3 (2 \lambda_k-1)^2}
	\\
		-	\frac{24 \pi  \left(2 g_k-\beta_g\right)}{g_k^2}+23
		=	0
	\,\,\text{,}
\end{multline}
\begin{multline}
		-	\frac{216 b_k^2 \left(9 g_k \beta_b-9 b_k \beta_g+72 b_k g_k+\beta_g-8 g_k\right)}{g_k (18 b_k+4 \lambda_k-3)^3}
	\\
		-	\frac{18 b_k \left(4 g_k \beta_b-4 b_k \beta_g+24 b_k g_k+\beta_g-6 g_k\right)}{g_k (18 b_k+4 \lambda_k-3)^2}
	\\
		-	\frac{2 (6 b_k \lambda_k-1) \left(-20 \beta_b+40 b_k+\frac{5 \beta_g}{g_k}-30\right)}{3 (2 \lambda_k-1)^2}
	\\
		-	\frac{-186 g_k \beta_b+186 b_k \beta_g-744 b_k g_k+29 \beta_g-116 g_k}{60 g_k (18 b_k+4 \lambda_k-3)}
	\\
		-	\frac{-40 b_k \beta_b+10 \beta_b+80 b_k^2-20 b_k+\frac{271 \beta_g}{36 g_k}-\frac{271}{9}}{1-2 \lambda_k}
		+	\frac{24 \pi  \left(g_k \beta_b-b_k \beta_g\right)}{g_k^2}
	\\
		-	\frac{\left(-72 b_k^2 \lambda_k+30 b_k \lambda_k+9 b_k-4\right) \left(80-\frac{10 \beta_g}{g_k}\right)}{9 (2 \lambda_k-1)^3}
		+	\frac{15 \left(4 g_k-\beta_g\right)}{2 g_k (2 \lambda_k-1)}
		-	\frac{872}{45} 
		=	0
	\,\,\text{,}
\end{multline}
\label{eq:beta2}%
\end{subequations}%
The $\beta$-functions \eqref{eq:beta2} are the main result of this appendix and underlie the analysis of the gravitational RG flow performed in section~\ref{sec:rg_structure}.

We remark that, recently, ref.\ \cite{Ohta:2015efa} derived a similar flow equation based on a physical gauge fixing condition. While the resulting  equation gives rise to a qualitatively similar structure in terms of fixed points, it also possess a singular hypersurface of codimension 1 that separates the NGFP from the classical region, see \cite{deBrito:2018jxt} for further analysis. Thus the corresponding solutions do not exhibit a crossover from the NGFP to a classical regime which is crucial for connecting the construction to a viable low-energy dynamics.


\begin{thebibliography}{99}
\bibitem{Riess:1998cb}
A.~G.~Riess {\it et al.} [Supernova Search Team],
Astron.\ J.\  {\bf 116} (1998) 1009,
astro-ph/9805201.

\bibitem{Perlmutter:1998np}
S.~Perlmutter {\it et al.} [Supernova Cosmology Project Collaboration],
Astrophys.\ J.\  {\bf 517} (1999) 565,
astro-ph/9812133.

\bibitem{Ade:2015lrj}
P.~A.~R.~Ade {\it et al.} [Planck Collaboration],
Astron.\ Astrophys.\  {\bf 594} (2016) A20,
arXiv:1502.02114 [astro-ph.CO].

\bibitem{Starobinsky:1980te}
  A.~A.~Starobinsky,
  Phys.\ Lett.\ B {\bf 91} (1980) 99.

  \bibitem{Mukhanov:1981xt}
  V.~F.~Mukhanov and G.~V.~Chibisov,
  JETP Lett.\  {\bf 33} (1981) 532
   [Pisma Zh.\ Eksp.\ Teor.\ Fiz.\  {\bf 33} (1981) 549].

  \bibitem{Starobinsky:1983zz}
  A.~A.~Starobinsky,
  Sov.\ Astron.\ Lett.\  {\bf 9} (1983) 302.


\bibitem{Niedermaier:2006wt}
M.~Niedermaier and M.~Reuter,
Living Rev.\ Rel.\  {\bf 9} (2006) 5.

\bibitem{Codello:2008vh}
A.~Codello, R.~Percacci and C.~Rahmede,
Annals Phys.\  {\bf 324} (2009) 414,
arXiv:0805.2909.

\bibitem{Litim:2011cp}
D.~F.~Litim,
Phil.\ Trans.\ Roy.\ Soc.\ Lond.\ A {\bf 369} (2011) 2759,
arXiv:1102.4624.

\bibitem{Reuter:2012id}
M.~Reuter and F.~Saueressig,
New J.\ Phys.\  {\bf 14} (2012) 055022,
arXiv:1202.2274.


\bibitem{Nagy:2012ef}
S.~Nagy,
Annals Phys.\  {\bf 350} (2014) 310,
arXiv:1211.4151.

\bibitem{Eichhorn:2017egq}
A.~Eichhorn,
arXiv:1709.03696 [gr-qc].

\bibitem{Percacci:2017fkn}
R.~Percacci,
``An Introduction to Covariant Quantum Gravity and Asymptotic Safety,''
World Scientific, 2017.


\bibitem{Bonanno:2001hi}
A.~Bonanno and M.~Reuter,
Phys.\ Lett.\ B {\bf 527} (2002) 9,
astro-ph/0106468.

\bibitem{Bonanno:2001xi}
A.~Bonanno and M.~Reuter,
Phys.\ Rev.\ D {\bf 65} (2002) 043508,
hep-th/0106133.

\bibitem{Guberina:2002wt}
B.~Guberina, R.~Horvat and H.~Stefancic,
Phys.\ Rev.\ D {\bf 67} (2003) 083001,
hep-ph/0211184.

\bibitem{Reuter:2003ca}
M.~Reuter and H.~Weyer,
Phys.\ Rev.\ D {\bf 69} (2004) 104022,
hep-th/0311196.

\bibitem{Babic:2004ev}
A.~Babic, B.~Guberina, R.~Horvat and H.~Stefancic,
Phys.\ Rev.\ D {\bf 71} (2005) 124041,
astro-ph/0407572.

\bibitem{Reuter:2005kb}
M.~Reuter and F.~Saueressig,
JCAP {\bf 0509} (2005) 012,
hep-th/0507167.

\bibitem{Bonanno:2005mt}
A.~Bonanno, G.~Esposito, C.~Rubano and P.~Scudellaro,
Class.\ Quant.\ Grav.\  {\bf 23} (2006) 3103,
astro-ph/0507670.

\bibitem{Bonanno:2007wg}
A.~Bonanno and M.~Reuter,
JCAP {\bf 0708} (2007) 024,
arXiv:0706.0174 [hep-th].

\bibitem{Weinberg:2009wa}
S.~Weinberg,
Phys.\ Rev.\ D {\bf 81} (2010) 083535,
arXiv:0911.3165 [hep-th].

\bibitem{Bonanno:2010bt}
A.~Bonanno, A.~Contillo and R.~Percacci,
Class.\ Quant.\ Grav.\  {\bf 28} (2011) 145026,
arXiv:1006.0192 [gr-qc].

\bibitem{Cai:2011kd}
Y.~F.~Cai and D.~A.~Easson,
Phys.\ Rev.\ D {\bf 84} (2011) 103502,
arXiv:1107.5815 [hep-th].

\bibitem{Contillo:2011ag}
A.~Contillo, M.~Hindmarsh and C.~Rahmede,
Phys.\ Rev.\ D {\bf 85} (2012) 043501,
arXiv:1108.0422 [gr-qc].

\bibitem{Cai:2013caa}
Y.~F.~Cai, Y.~C.~Chang, P.~Chen, D.~A.~Easson and T.~Qiu,
Phys.\ Rev.\ D {\bf 88} (2013) 083508,
arXiv:1304.6938.

\bibitem{Copeland:2013vva}
E.~J.~Copeland, C.~Rahmede and I.~D.~Saltas,
Phys.\ Rev.\ D {\bf 91} (2015) 103530,
arXiv:1311.0881 [gr-qc].

\bibitem{Saltas:2015vsc}
I.~D.~Saltas,
JCAP {\bf 02} (2016) 048,
arXiv:1512.06134 [hep-th].

\bibitem{Tronconi:2017wps}
A.~Tronconi,
JCAP {\bf 07} (2017) 015,
arXiv:1704.05312 [gr-qc].



\bibitem{Bonanno:2012jy}
A.~Bonanno,
Phys.\ Rev.\ D {\bf 85} (2012) 081503,
arXiv:1203.1962 [hep-th].

\bibitem{Bonanno:2015fga}
A.~Bonanno and A.~Platania,
Phys.\ Lett.\ B {\bf 750} (2015) 638,
arXiv:1507.03375 [gr-qc].

\bibitem{Bonanno:2018gck}
A.~Bonanno, A.~Platania and F.~Saueressig,
arXiv:1803.02355 [gr-qc].

\bibitem{DOdorico:2015jtl}
G.~D'Odorico and F.~Saueressig,
Phys.\ Rev.\ D {\bf 92} (2015) 124068,
arXiv:1511.00247 [gr-qc].


\bibitem{Bonanno:2017pkg}
A.~Bonanno and F.~Saueressig,
Comptes Rendus Physique {\bf 18} (2017) 254,
arXiv:1702.04137 [hep-th].

\bibitem{Weinberg:1980gg}
  S.~Weinberg
  in \textit{General Relativity, an Einstein Centenary Survey},
  S.W.~Hawking and W.~Israel (Eds.),
  Cambridge University Press, 1979.

\bibitem{Reuter:1996cp}
M.~Reuter,
Phys.\ Rev.\ D {\bf 57} (1998) 971,
hep-th/9605030.
  
\bibitem{Gies:2016con}
H.~Gies, B.~Knorr, S.~Lippoldt and F.~Saueressig,
Phys.\ Rev.\ Lett.\  {\bf 116} (2016) 211302,
arXiv:1601.01800.

\bibitem{Manrique:2011jc}
E.~Manrique, S.~Rechenberger and F.~Saueressig,
Phys.\ Rev.\ Lett.\  {\bf 106} (2011) 251302,
arXiv:1102.5012 [hep-th].

\bibitem{Liu:2018aa}
L.~H.~Liu, T.~Prokopec and A.~A.~Starobinsky, Phys.Rev. D {\bf 98} (2018)  043505,
arXiv:1806.05407 [gr-qc].

\bibitem{Dietz:2012ic}
J.~A.~Dietz and T.~R.~Morris,
JHEP {\bf 01} (2013) 108,
arXiv:1211.0955.

\bibitem{Demmel:2015oqa}
M.~Demmel, F.~Saueressig and O.~Zanusso,
JHEP {\bf 08} (2015) 113,
arXiv:1504.07656.

\bibitem{Machado:2007ea} 
  P.~F.~Machado and F.~Saueressig,
  Phys.\ Rev.\ D {\bf 77}, 124045 (2008),
  arXiv:0712.0445 [hep-th].



\bibitem{Martin:2013tda}
  J.~Martin, C.~Ringeval and V.~Vennin,
  Phys.\ Dark Univ.\  {\bf 5-6} (2014) 75,
  arXiv:1303.3787 [astro-ph.CO].

\bibitem{Bezrukov:2007ep}
  F.~L.~Bezrukov and M.~Shaposhnikov,
  Phys.\ Lett.\ B {\bf 659} (2008) 703,
  arXiv:0710.3755 [hep-th].
  
  \bibitem{Barvinsky:2008ia}
  A.~O.~Barvinsky, A.~Y.~Kamenshchik and A.~A.~Starobinsky,
  JCAP {\bf 0811} (2008) 021,
  arXiv:0809.2104 [hep-ph].
  
\bibitem{Netto:2015cba}
T.~d.~P.~Netto, A.~M.~Pelinson, I.~L.~Shapiro and A.~A.~Starobinsky,
Eur.\ Phys.\ J.\ C {\bf 76} (2016) 544,
arXiv:1509.08882 [hep-th].
  
\bibitem{Ade:2015xua}
  P.~A.~R.~Ade {\it et al.} [Planck Collaboration],
  Astron.\ Astrophys.\  {\bf 594} (2016) A13,
  arXiv:1502.01589 [astro-ph.CO].

\bibitem{Patrignani:2016xqp}
  C.~Patrignani {\it et al.} [Particle Data Group],
  Chin.\ Phys.\ C {\bf 40} (2016) 100001.

\bibitem{Riess:2016jrr}
  A.~G.~Riess {\it et al.},
  Astrophys.\ J.\  {\bf 826} (2016) 56,
  arXiv:1604.01424 [astro-ph.CO].

  \bibitem{Ohta:2015efa}
  N.~Ohta, R.~Percacci and G.~P.~Vacca,
  Phys.\ Rev.\ D {\bf 92} (2015) 061501,
  arXiv:1507.00968 [hep-th].
  
\bibitem{deBrito:2018jxt}
G.~P.~De Brito, N.~Ohta, A.~D.~Pereira, A.~A.~Tomaz and M.~Yamada,
arXiv:1805.09656 [hep-th].
  
\bibitem{Rechenberger:2012pm}
S.~Rechenberger and F.~Saueressig,
Phys.\ Rev.\ D {\bf 86} (2012) 024018,
arXiv:1206.0657 [hep-th].
  
\bibitem{Lauscher:2002sq}
O.~Lauscher and M.~Reuter,
Phys.\ Rev.\ D {\bf 66} (2002) 025026,
hep-th/0205062.
  
\bibitem{Codello:2007bd}
A.~Codello, R.~Percacci and C.~Rahmede,
Int.\ J.\ Mod.\ Phys.\ A {\bf 23} (2008) 143,
arXiv:0705.1769 [hep-th].

\bibitem{Falls:2014tra}
K.~Falls, D.~F.~Litim, K.~Nikolakopoulos and C.~Rahmede,
Phys.\ Rev.\ D {\bf 93} (2016) 104022,
arXiv:1410.4815 [hep-th].

\bibitem{Reuter:2001ag}
  M.~Reuter and F.~Saueressig,
  Phys.\ Rev.\ D {\bf 65} (2002) 065016,
  hep-th/0110054.

\bibitem{Reuter:2004nx}
M.~Reuter and H.~Weyer,
JCAP {\bf 0412} (2004) 001,
hep-th/0410119.

\bibitem{Amelino-Camelia:2015dqa}
  G.~Amelino-Camelia, M.~Arzano, G.~Gubitosi and J.~Magueijo,
  Int.\ J.\ Mod.\ Phys.\ D {\bf 24} (2015) no.12,  1543002,
  arXiv:1505.04649 [gr-qc].
  
\bibitem{Brighenti:2016xng}
  F.~Brighenti, G.~Gubitosi and J.~Magueijo,
  Phys.\ Rev.\ D {\bf 95} (2017)  063534,
  arXiv:1612.06378 [gr-qc].
  
\bibitem{Codello:2015pga}
A.~Codello and R.~K.~Jain,
Class.\ Quant.\ Grav.\  {\bf 34} (2017) 035015,
arXiv:1507.07829 [astro-ph.CO].

\bibitem{Codello:2016neo}
A.~Codello and R.~K.~Jain,
Eur.\ Phys.\ J.\ C {\bf 78} (2018) 357,
arXiv:1603.00028 [gr-qc].
  
\bibitem{Dona:2013qba}
P.~Don\`{a}, A.~Eichhorn and R.~Percacci,
Phys.\ Rev.\ D {\bf 89} (2014) 084035,
arXiv:1311.2898 [hep-th].

\bibitem{Christiansen:2017cxa}
N.~Christiansen, D.~F.~Litim, J.~M.~Pawlowski and M.~Reichert,
Phys.\ Rev.\ D {\bf 97} (2018) 106012,
arXiv:1710.04669 [hep-th].

\bibitem{Alkofer:2018fxj}
N.~Alkofer and F.~Saueressig, Annals Phys.\ {\bf 396} (2018) 173,
arXiv:1802.00498 [hep-th].
  
\bibitem{Martin:2012bt}
J.~Martin,
Comptes Rendus Physique {\bf 13} (2012) 566,
arXiv:1205.3365 [astro-ph.CO].

\bibitem{Sola:2013gha}
J.~Sola,
J.\ Phys.\ Conf.\ Ser.\  {\bf 453} (2013) 012015,
arXiv:1306.1527 [hep-th].

\bibitem{Padilla:2015aaa}
A.~Padilla,
arXiv:1502.05296 [hep-th].

\bibitem{Gies:2014xha}
H.~Gies and R.~Sondenheimer,
Eur.\ Phys.\ J.\ C {\bf 75} (2015) 68,
arXiv:1407.8124 [hep-ph].

\bibitem{Eichhorn:2015kea}
A.~Eichhorn, H.~Gies, J.~Jaeckel, T.~Plehn, M.~M.~Scherer and R.~Sondenheimer,
JHEP {\bf 1504} (2015) 022,
arXiv:1501.02812 [hep-ph].

\bibitem{Eichhorn:2017als}
A.~Eichhorn, Y.~Hamada, J.~Lumma and M.~Yamada,
Phys.\ Rev.\ D {\bf 97} (2018) 086004,
arXiv:1712.00319 [hep-th].

\bibitem{Ruf:2017xon}
M.~S.~Ruf and C.~F.~Steinwachs,
Phys.\ Rev.\ D {\bf 97} (2018) 044050,
arXiv:1711.07486 [gr-qc].

\bibitem{Ohta:2017trn}
N.~Ohta,
PTEP {\bf 2018} (2018) 033B02,
arXiv:1712.05175 [hep-th].
  
\bibitem{Chiba:2003ir}
  T.~Chiba,
  Phys.\ Lett.\ B {\bf 575} (2003) 1,
  astro-ph/0307338.


\bibitem{Wetterich:1992yh}
C.~Wetterich,
Phys.\ Lett.\ B {\bf 301} (1993) 90,
arXiv:1710.05815.

\bibitem{Morris:1993qb}
T.~R.~Morris,
Int.\ J.\ Mod.\ Phys.\ A {\bf 9} (1994) 2411,
hep-ph/9308265.

\bibitem{Reuter:1993kw}
M.~Reuter and C.~Wetterich,
Nucl.\ Phys.\ B {\bf 417} (1994) 181.

\bibitem{Benedetti:2012dx}
D.~Benedetti and F.~Caravelli,
JHEP {\bf 1206} (2012) 017
Erratum: [JHEP {\bf 1210} (2012) 157],
arXiv:1204.3541 [hep-th].




\bibitem{Dietz:2013sba}
J.~A.~Dietz and T.~R.~Morris,
JHEP {\bf 07} (2013) 064,
arXiv:1306.1223.

\bibitem{Demmel:2014hla}
M.~Demmel, F.~Saueressig and O.~Zanusso,
Annals Phys.\  {\bf 359} (2015) 141,
arXiv:1412.7207.



\bibitem{Ohta:2015fcu}
N.~Ohta, R.~Percacci and G.~P.~Vacca,
Eur.\ Phys.\ J.\ C {\bf 76} (2016) 46,
arXiv:1511.09393 [hep-th].
  
\end{thebibliography}
\end{document}